\documentclass[conference]{IEEEtran}
\IEEEoverridecommandlockouts

\usepackage{cite}
\usepackage{multirow}
\usepackage{booktabs}
\usepackage[table]{xcolor}
\usepackage{colortbl}
\usepackage{array}
\usepackage{amsmath,amssymb,amsfonts}
\usepackage{algorithmic}
\usepackage{graphicx}
\usepackage{textcomp}
\usepackage{orcidlink}
\usepackage{paralist}
\usepackage{listings}
\usepackage[most]{tcolorbox}

\newtcolorbox{mybox}[2][]{%
    enhanced,
    arc=3pt,
    fonttitle=\small,
    outer arc=3pt,
    boxrule=0.4pt,
    colback=white,
    title=#2,
    colframe=black,
    toprule=0.8mm,
    bottomrule=0.4pt,
    top=0pt,
    bottom=0pt,
    left=5pt,
    right=5pt,
    #1 
}

\definecolor{lightgray}{rgb}{0.95,0.95,0.95}
\definecolor{lightgreen}{rgb}{0.8, 1, 0.8}
\definecolor{lightred}{rgb}{1, 0.8, 0.8}
\definecolor{lightyellow}{rgb}{1, 1, 0.8}

\def\BibTeX{{\rm B\kern-.05em{\sc i\kern-.025em b}\kern-.08em
    T\kern-.1667em\lower.7ex\hbox{E}\kern-.125emX}}

\begin{document}

\title{Generating Software Architecture Description from Source Code using Reverse Engineering and Large Language Model}

\author{
    \IEEEauthorblockN{
        1\textsuperscript{st} Ahmad Hatahet \IEEEauthorrefmark{1},
        2\textsuperscript{nd} Christoph Knieke \IEEEauthorrefmark{2},
        3\textsuperscript{rd} Andreas Rausch \IEEEauthorrefmark{3},
    }
    \IEEEauthorblockA{
        \textit{Institute for Software and Systems Engineering} \\
        \textit{Technical University of Clausthal}\\
        Clausthal-Zellerfeld, Germany \\
        \IEEEauthorrefmark{1}ahmad.hatahet@tu-clausthal.de \orcidlink{0009-0009-7677-7514},
        \IEEEauthorrefmark{2}christoph.knieke@tu-clausthal.de \orcidlink{0009-0006-8018-2351},
        \IEEEauthorrefmark{3}andreas.rausch@tu-clausthal.de \orcidlink{0000-0002-6850-6409}
    }
}

\maketitle

\begin{abstract}

Software Architecture Descriptions (SADs) are essential for managing the inherent complexity of modern software systems. They enable high-level architectural reasoning, guide design decisions, and facilitate effective communication among diverse stakeholders. However, in practice, SADs are often missing, outdated, or poorly aligned with the system’s actual implementation. Consequently, developers are compelled to derive architectural insights directly from source code—a time-intensive process that increases cognitive load, slows new developer onboarding, and contributes to the gradual degradation of clarity over the system’s lifetime. To address these issues, we propose a semi-automated generation of SADs from source code by integrating reverse engineering (RE) techniques with a Large Language Model (LLM). Our approach recovers both static and behavioral architectural views by extracting a comprehensive component diagram, filtering architecturally significant elements (core components) via prompt engineering, and generating state machine diagrams to model component behavior based on underlying code logic with few-shots prompting. This resulting views representation offer a scalable and maintainable alternative to traditional manual architectural documentation. This methodology, demonstrated using C++ examples, highlights the potent capability of LLMs to: 1) abstract the component diagram, thereby reducing the reliance on human expert involvement, and 2) accurately represent complex software behaviors, especially when enriched with domain-specific knowledge through few-shot prompting. These findings suggest a viable path toward significantly reducing manual effort while enhancing system understanding and long-term maintainability.

\end{abstract}


\section{Introduction} \label{intro}

Software Architecture (SA) is central to managing complexity during software development by establishing the system’s core structure, key components, and their interactions. As a foundational artifact, it guides long-term system evolution, supports informed technical decisions, and ensures alignment between system requirements and implementation. To effectively fulfill this role, the architecture must be explicitly defined and systematically documented.

Software Architecture Description (SAD) offers this necessary systematic formalization \cite{architecture_foundation}, capturing architectural knowledge in the form of structured views and artifacts. These views facilitate communication among stakeholders, from managers to developers, each requiring a tailored level of abstraction. In this context, two views are particularly critical: the static view, generally presented as component diagrams, which describes structural elements and their relationships; and the behavioral view, typically modeled through state machine diagrams, which captures component dynamics and operational lifecycles.

However, SADs are often missing, incomplete, or partially or severely outdated \cite{10.1007/978-3-642-39031-9_7,satish2016software,itsdartWhatConsequences}. Development teams frequently rely on source code as the sole reference point for understanding system structure and behavior. Yet, source code inherently lacks the necessary architectural abstraction and is thus not suited for broad stakeholder comprehension. This gap results in increased onboarding effort, reduced design clarity, and a gradual erosion of system-level understanding, particularly in real-world and complex projects \cite{itsdartWhatConsequences,rukmono2025expert}.

To reduce this deficit, reverse engineering (RE) \cite{reverse_engineer} has been extensively investigated as a method to recover architectural information directly from source code. While these techniques are useful for extracting structural representations, they often yield an overwhelming volume of low-level details or miss semantically significant connections. Furthermore, most solutions are limited to static views, leaving the crucial behavioral aspects of the architecture undocumented and unexamined \cite{Nicolaescu:808671}.

Recent advancements in Large Language Models (LLMs) present a significant new opportunity. With their strong capabilities in code comprehension and natural language generation \cite{10.1145/3597503.3639187}, LLMs can support a range of development tasks, including logic summarization, behavior interpretation, and visual diagram generation using specialized languages like PlantUML \cite{linkedin_plantuml,medium_plantuml}. Their capacity to generalize across languages and contexts makes them ideally suited for behavior modeling. Nevertheless, their effectiveness relies heavily on precise instructions and relevant contextual data, such as specific target code scripts, to avoid irregular behavioral interpretations.

In this work, we propose a hybrid approach that integrates RE and LLM reasoning to automate the generation of software architecture descriptions from source code. The process begins with RE to extract a comprehensive class diagram reflecting the system’s structural details. This low-level representation then requires abstraction into a higher-level component diagram—where a component encapsulates a cohesive group of classes and exposes defined interfaces. Since this abstraction step is crucial, project-specific, and context-dependent, it is delegated to the LLM, whose dynamic ability to interpret code semantics makes it ideally suited for reorganizing the structure into meaningful architectural components \cite{code_semmantic,hou2024large}. Once these core components are established, the LLM analyzes their internal logic and generates corresponding state machine diagrams to capture behavioral patterns. The overall approach is designed to maintain semantic accuracy while effectively managing input size constraints.

To address this gap, this paper investigates two core questions: \textbf{RQ1)} To what extent can SAD be automatically recovered from source code alone, structurally and behaviorally, without reliance on prior documentation or annotations? \textbf{RQ2)} Can an LLM, with carefully scoped input, generate accurate behavioral views (state machine diagrams) that accurately reflect the internal logic of architectural components? Through this investigation, we aim to demonstrate a practical solution for SAD generation that supports superior system comprehension and transparency.

The remainder of this paper is structured as follows. Section \ref{related} reviews the related work, Section \ref{approach} details our approach and methodology, Section \ref{result} presents our results and includes a thorough discussion. Finally, conclusions and future research directions are summarized in Section \ref{conclusion}.

\textit{\textbf{Note:} All diagram images, python notebooks, and prompts are available in our GitHub repository \cite{gh_repo}.}


\section{Related Work} \label{related}

Learning a software architecture from code involves the task of identifying and recovering the underlying architecture of a software system from its existing code base. The main objective of this process is to create a high-level architectural model, including its components, relationships, and the architectural patterns used in the implementation.

Reverse engineering architectural models from code is a crucial process \cite{reverse_engineer} because it provides a better understanding of a system’s structure and the patterns applied. It serves as a valuable source for future decision-making related to maintaining and further developing the system. Manually maintained documentation and models of a system tend to diverge from the actual source code of the system over time, reducing the reliability of such artifacts \cite{itsdartWhatConsequences}. In some cases, these documents may not even exist at all.

Static approaches focus on analyzing the source code of a software system without executing it \cite{73}.
These techniques use various code analysis tools to extract information from the source code, such as parsing source code to extract syntax and control flow information or using metrics to quantify certain properties of the code \cite{45}.

Various tools have attempted to automate architectural recovery, though with notable trade-offs. Tools like Doxygen \cite{doxygen} and SmartDraw \cite{smart_draw} generate component diagrams directly from source code. However, these diagrams often omit critical architectural details—such as class attributes, method implementations, and inter-class associations—rendering them inadequate for comprehensive static views. SmartDraw, in particular, captures associations inconsistently, limiting its reliability.

CLANG-UML \cite{clang} improves automation by generating diagrams in PlantUML format \cite{plantuml}, yet its output is often too granular. For real-world systems, the resulting diagrams are overwhelmed with low-level implementation details, complicating architectural reasoning. Similarly, Enterprise Architect (EA) \cite{ea} supports multi-language RE and rich modeling capabilities, but like its counterparts, struggles to deliver the necessary level of abstraction for architectural-level insight.

To address these gaps, several research efforts propose more tailored or hybrid solutions. Schiewe et al. \cite{73} introduce a language-agnostic methodology for component identification using an intermediate representation—LAAST—guided by heuristic rules. This approach abstracts away syntactic language features, aiming to detect higher-level architectural components. Walker et al. \cite{85}, targeting microservice systems, combine static analysis with runtime tracing to reconstruct distributed architectures. This hybrid approach effectively captures both structural and behavioral aspects of the system.

Some efforts are context-specific. Amalfitano et al. \cite{sar} propose a domain-tailored solution for embedded systems using C. Their custom tool automates the generation of software architecture documentation, producing package, component, component and connector (C\&C), and state machine diagrams. This approach is particularly noteworthy because its success stems from effectively leveraging specific code characteristics inherent to their embedded system use case. For instance, the tool defines the package diagram based on the system's folder structure, and it extracts system states and their triggers from C code references to switch-based task tables. This illustrates how domain-specific knowledge of code organization and patterns can be instrumental in successful architecture recovery.

Other recent efforts pursue automation through graph-based techniques. Zhang et al. \cite{fusion} merge dependency graphs with directory structures, assigning edge weights and applying clustering to extract architectural components. Puchala et al. \cite{ensemble} follow a similar clustering-based methodology, while Sajji et al. \cite{gnn_1} adopt Graph Neural Networks (GNNs) to learn abstraction from structural code representations.

Overall, while numerous tools and methods exist, current techniques often trade off between abstraction and completeness. Effective architecture recovery remains challenging, particularly in ensuring scalability, accuracy, and clarity across diverse codebases and application domains.

\section{Approach and Methodology} \label{approach}

This work presents a semi-automated approach for generating software architecture descriptions from source code, focusing on both the static and behavioral views.
The process begins by analyzing the source code via RE, producing an initial detailed class diagram.
This diagram is translated into PlantUML format using a direct mapping method that captures the class with its attributes and operations, and general associations between the classes.

The LLM receives the full plain PlantUML then it identifies and retains only the architecturally significant classes essential for understanding the system using prompt engineering, we refer to them as core components.
The output of the LLM is a list of the core components, which help refine and abstract the class diagram into a component diagram.

For the behavioral view, we generate a state machine diagram for each core component. This is done by providing its corresponding source code to the LLM along with solution plan and few-shot example to infer possible states and transitions based on internal logic and method behaviors.

The result is a dual-view SAD: 1) an abstract component diagram capturing the system’s structure and 2) a set of state machine diagrams illustrating key component behaviors.
This methodology balances automation with guided abstraction to support system understanding and architectural clarity.

\begin{figure}[!h]

\begin{center}
    \includegraphics[width=\columnwidth]{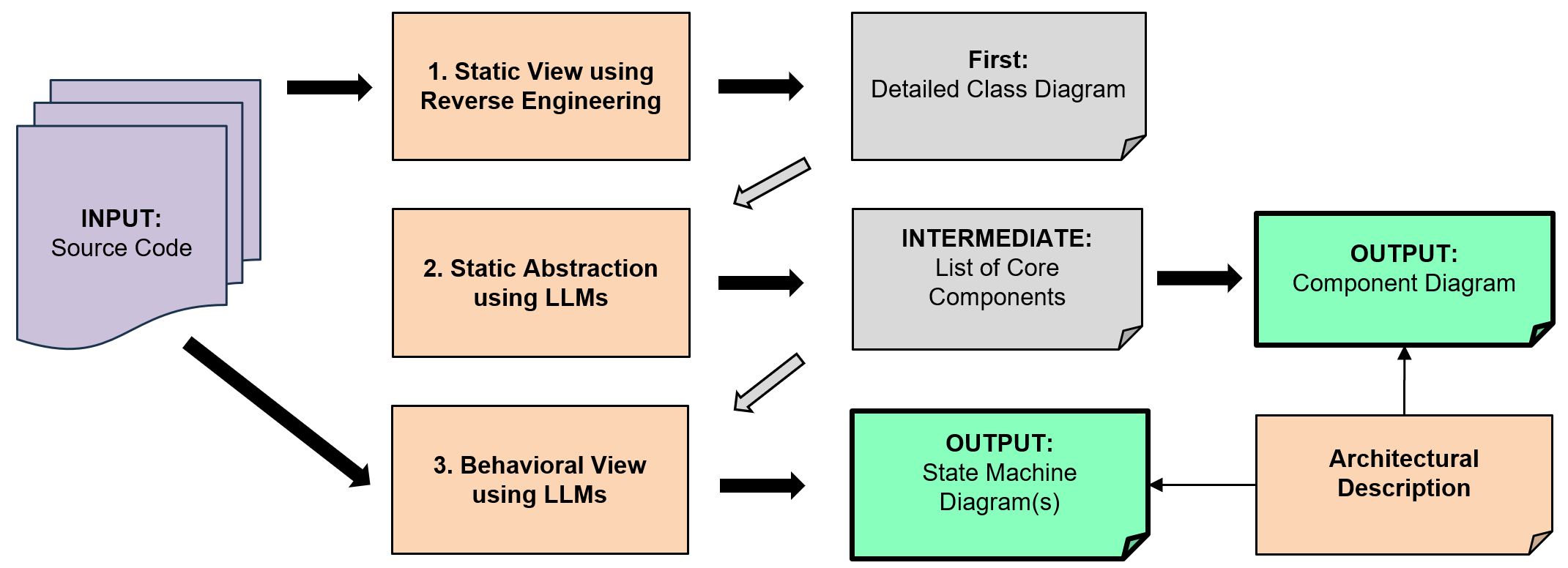}
\end{center}
\caption{Proposed approach: the source code is used as an input then after RE, a detailed class diagram generated, after using LLM to abstract it to a component diagram, the LLM also generate multiple state machine diagrams for each component. Component and stat machine diagrams are the output forming the final SAD.}
\label{fig:main_approach}

\end{figure}

\subsection{Static View using Reverse Engineering}

The process begins with extracting a detailed structural representation of the system using Enterprise Architect (EA) \cite{ea}.
The source code is imported into EA, which statically analyzes the system and generates a comprehensive class diagram.
This diagram reads all the classes and their interconnections, forming an exhaustive map of the system’s structure, and often includes low-level elements that may obscure the overall architecture.
As such, it serves as an initial artifact from which a more meaningful abstraction will be derived in the subsequent step.

\subsection{Static Abstraction using LLM}

To address this, EA \cite{ea} exports the component diagram to an XML file.
Afterward, a custom Python function parses this XML file into PlantUML format, enabling easier manipulation.
An example PlantUML snippet is shown below:

\begin{mybox}[colframe=teal!75!black]{PlantUML Example of only two Classes}
\begin{lstlisting}[breaklines=true, columns=fullflexible]
class CoffeeMachine {
    -waterLevel
    +startBrew()
}
class Boiler {
    -temperature
    +heatWater()
}
CoffeeMachine -- Boiler
\end{lstlisting}
\end{mybox}

The full PlantUML code is then passed to GPT-4o version 2024-11-20 \cite{gpt4o}, which identifies architecturally significant classes—referred to as core components—using a system prompt. A second script filters the detailed component diagram in the XML file, retaining only these core components and their relationships to form an abstract structural view.

\subsection{Behavioral View using LLM}

For behavioral modeling, the source code of each core component is submitted to GPT-4o version 2024-11-20 \cite{gpt4o} alongside few-shot prompting \cite{few_shots}. The examples teach the LLM the link between code and state machine diagrams. This guides the model in generating state machine diagrams that reflect internal logic and transitions.

Together, these steps produce a refined, dual-view architecture description: a simplified component diagram for structure and targeted state machine diagrams for behavior. The integration balances automation with selective abstraction, enhancing clarity and architectural relevance. Finally, the evaluation of the generated diagrams is compared with ground truth diagrams to validate the effectiveness of the approach and the quality of the diagrams.


\section{Results and Discussion} \label{result}

This section presents and analyzes the outcomes of our approach for generating software architecture descriptions from source code, focusing on both the static and behavioral views. The static view is captured in the form of component diagrams, while the behavioral view is expressed through state machine diagrams. We evaluate the quality and fidelity of the generated architectural artifacts against ground truth diagrams using clearly defined criteria and through comparative inspection.


\subsection{Static View: Component Diagram Extraction}

To ensure a valid evaluation, our starting point involved selecting systems with available ground truth representations of both class and state machine diagrams. Given the industrial relevance and our collaboration with IAV\footnote{https://www.iav.com/}, the source code had to be in C++. Among available tools, IBM Rhapsody\footnote{https://www.ibm.com/support/pages/ibm-rational-rhapsody-84} version 8.4 provided a curated set of C++ examples that included architectural documentation. From this set, we selected two illustrative systems: the \textit{Coffee Machine} and the \textit{Dishwasher}.

\begin{figure}[!htp]
    \centering
    \includegraphics[width=\columnwidth]{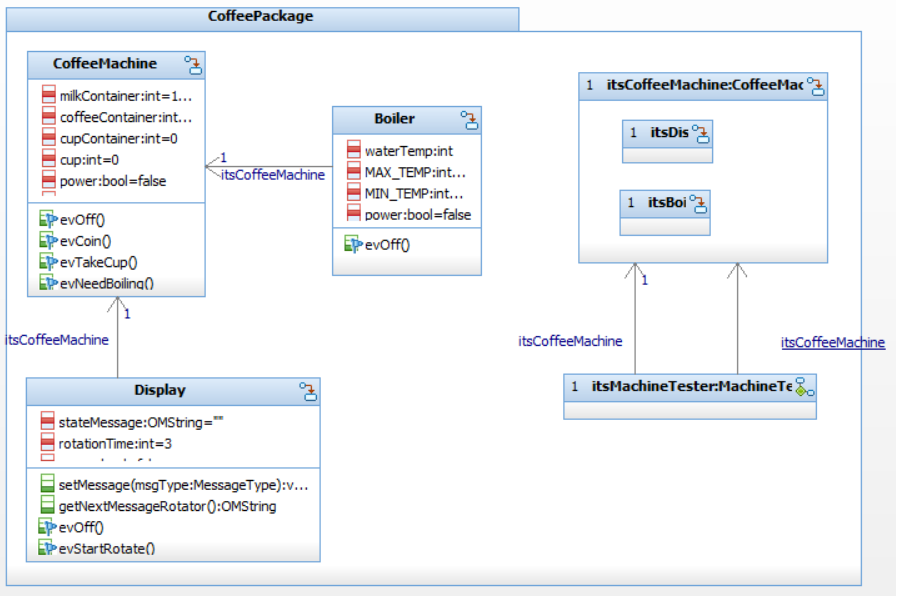}
    \caption{\centering Ground truth component diagram from Rhapsody showing the core components of the Coffee Machine}
    \label{fig:coffee_class}
\end{figure}

\begin{figure}[!htp]
    \centering
    \includegraphics[width=\columnwidth]{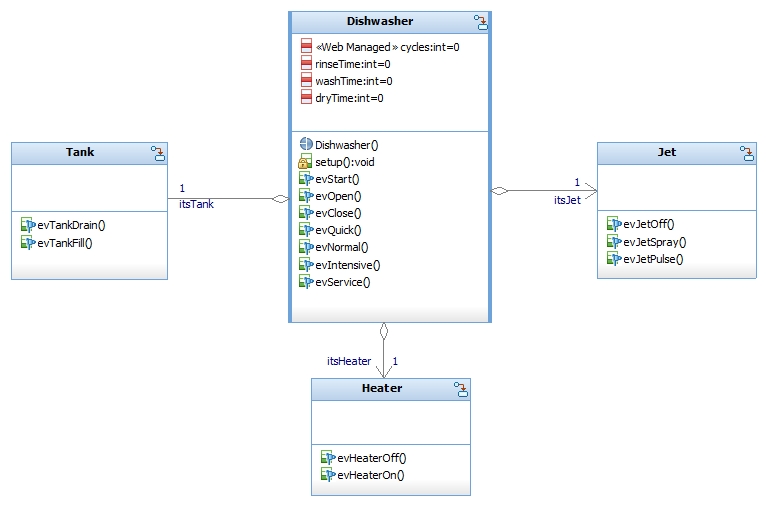}
    \caption{\centering Ground truth component diagram from Rhapsody showing the core components of the Dishwasher}
    \label{fig:dishwasher_class}
\end{figure}

\begin{figure}[!h]
    \centering
    \includegraphics[width=\columnwidth, height=8cm]{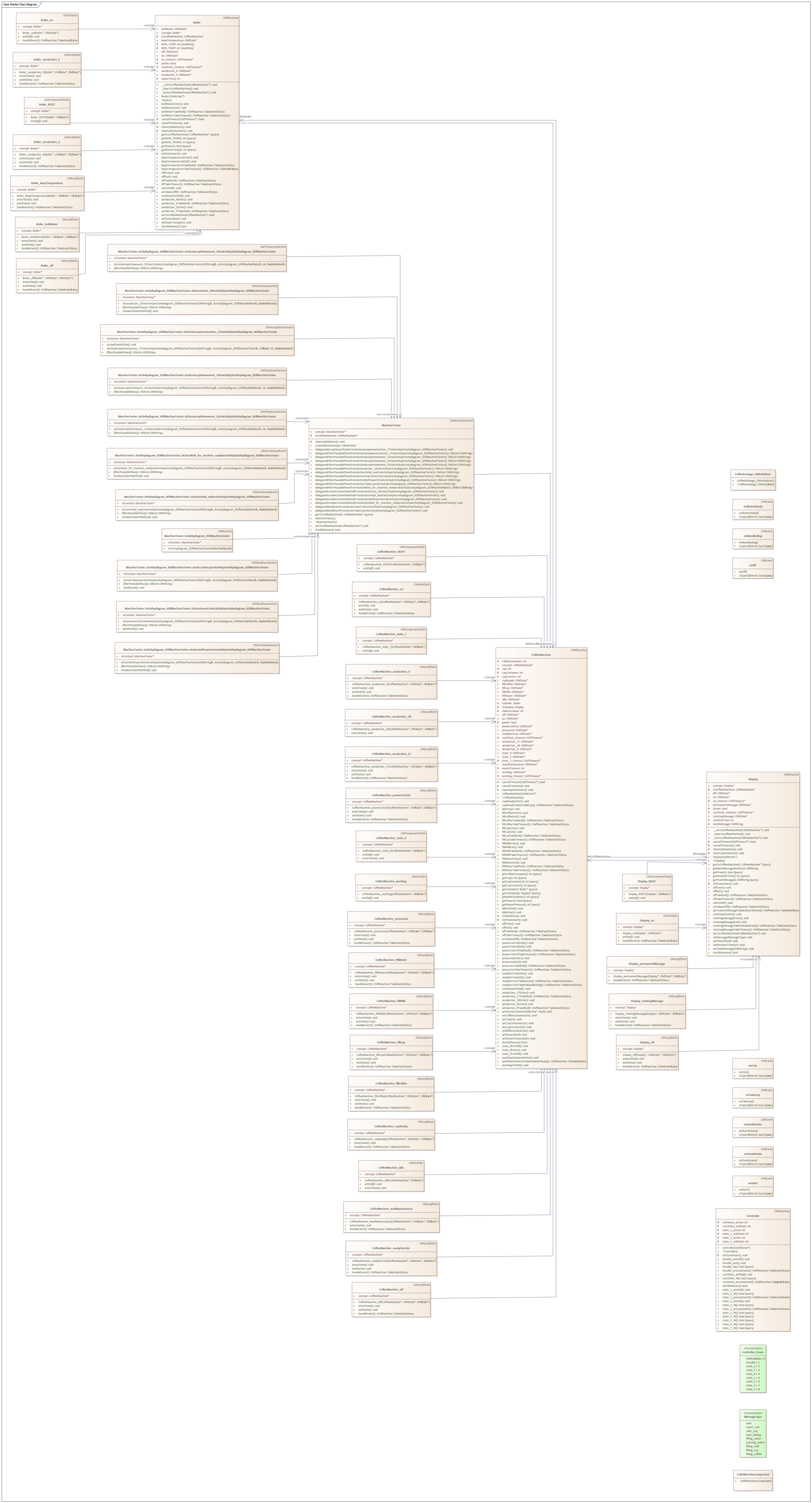}
    \caption{\centering Overview of the complexity of all Coffee Machine's classes after RE the source code via EA (full resolution available in our repository: \emph{diagrams/CofeeMachine/class\_diagram/class-diagram.png})}
    \label{fig:reverse_overview}
\end{figure}

The reference component diagrams of these systems shown in Fig.~\ref{fig:coffee_class} and Fig.~\ref{fig:dishwasher_class} respectively.
We reverse-engineered the corresponding C++ code using EA \cite{ea}, which resulted in highly detailed and extensive component diagrams as depicted in Fig.~\ref{fig:reverse_overview}.
These diagrams, while accurate, included numerous auxiliary classes that obscure the essential structural components. Therefore, further abstraction was necessary.

\subsection{Core Component Identification and Abstraction}

As described, to facilitate automated reasoning with LLMs, we implemented a naïve but deterministic translation mechanism to convert the EA \cite{ea} component diagrams into PlantUML syntax \cite{plantuml}.
The translation included all class names, their attributes, operations, and direct associations, preserving the structure observed in the EA \cite{ea} generated diagrams.

This PlantUML specification was then passed to GPT-4o version 2024-11-20 \cite{gpt4o} with a prompt designed to extract the core architectural components.

\begin{mybox}[colframe=violet!75!black]{Find Core Components Prompt}
\begin{lstlisting}[basicstyle=\scriptsize, breaklines=true, columns=fullflexible, breakindent=0pt]
<role>
You are an expert software engineer.
</role>
<goal>
Extract an abstract view from the classes keeping only the most important classes.
</goal>

<description>
You will receive a component diagram in PlantUML format highlighting all classes and their attributes and operations.
Additionally, the relationships between classes are also included in one form of association which is not influential but helps you know which class connected to which class.
</description>

<plantuml>
{plantuml_text}
</plantuml>
\end{lstlisting}
\end{mybox}

To evaluate the abstracted component diagram, a number of questions are proposed to systematically assess it:

\begin{enumerate}
    \item[Q1)] Did the LLM identified the correct components to represent the system at an appropriate level of abstraction?
    \item[Q2)] Are the relationships and interfaces between components correctly depicted, reflecting the system's intended interactions?
    \item[Q3)] Are the components and interactions correctly labeled?
\end{enumerate}

Remarkably, the LLM demonstrated the ability to abstract away auxiliary utility classes and identify structurally and functionally central elements for both examples.

For the Coffee Machine, GPT-4o \cite{gpt4o} selected: \textit{"CoffeeMachine"}, \textit{"Boiler"}, \textit{"Display"}, \textit{"MachineTester"}, and \textit{"Controller"}.
Although \textit{"MachineTester"} and \textit{"Controller"} were not in the ground truth, upon inspection their inclusion was justifiable due to their functional relevance.
Auxiliary classes like \textit{"Boiler\_boilWater"} were correctly excluded.
A similar abstraction quality was observed for the Dishwasher example, where the model returned \textit{"AbstractFactory"}, \textit{"Dishwasher"}, \textit{"Heater"}, \textit{"Jet"}, and \textit{"Tank"}. Only \textit{"AbstractFactory"} was irrelative to the ground truth.

These results answer \textit{Q1} and confirm the LLM’s potential for filtering architectural noise and isolating conceptually meaningful components based solely on structural input.

Answering \textit{Q2}: we can see in Fig.~\ref{fig:coffe_state_gt} and Fig.~\ref{fig:dish_state_gt}, the connection between the components are clear and understandable.
However, the components \textit{"Controller"} and \textit{"AbstractFactory"} are isolated. This presentation is the result of the RE when we imported the source code into EA \cite{ea}. The RE did not find any connection from those two components to the other ones rendering those two to be isolated.

The LLM has not changed the names of the core components when it has selected them from the PlantUML text.
Therefore, the component names came from the RE which is a reflection of the source code.
Consequently, the relationship labels are also not impacted from the LLM, however, the RE did not call them with clear labels (i.e. \textit{"itsDisplay"} or \textit{"itsHeater"}).
Answering \textit{Q3}: The component names are understandable, but the connection labels are not.

The overall resulted component diagram is indeed an abstraction from the class diagram, which the LLM did an acceptable job with.
The remaining factors are dependent on the RE process, which may be improved in the future.


\begin{figure}[!h]
    \centering
    \includegraphics[scale=0.25]{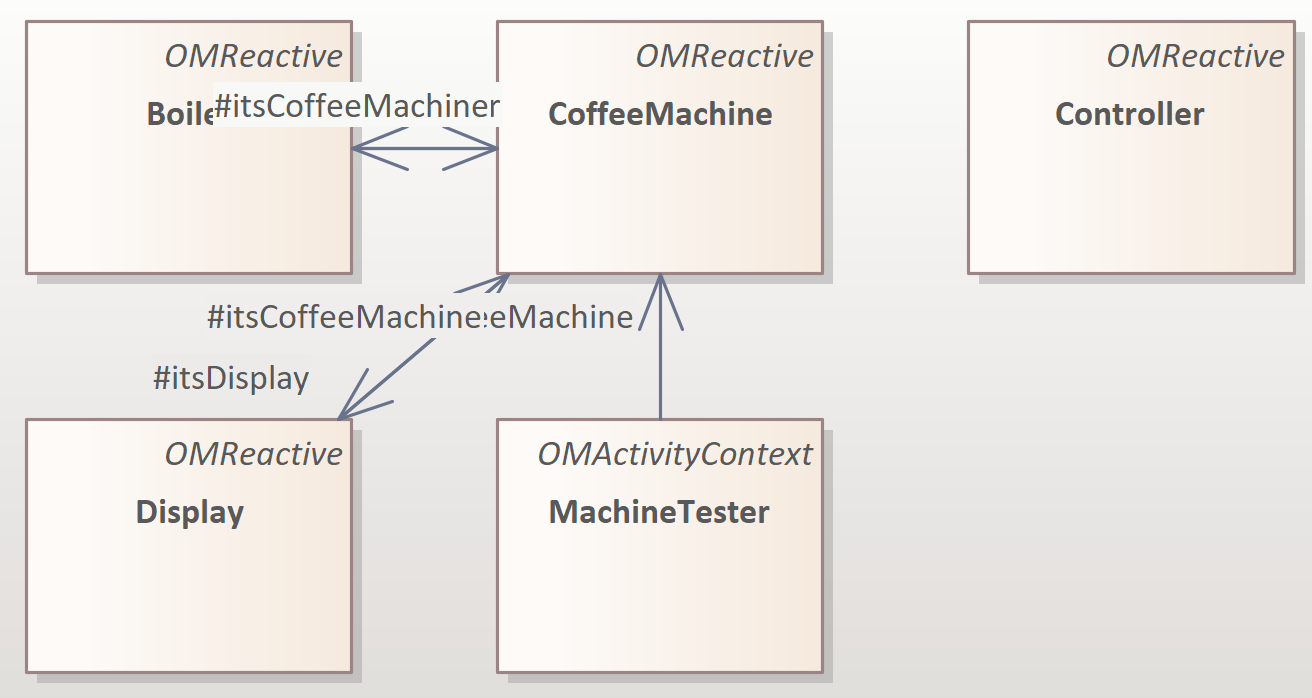}
    \caption{\centering Static View - Component Diagram of the Coffee Machine core components after abstraction using LLM}
    \label{fig:coffe_class_ea}
\end{figure}

\begin{figure}[!h]
    \centering
    \includegraphics[scale=0.6]{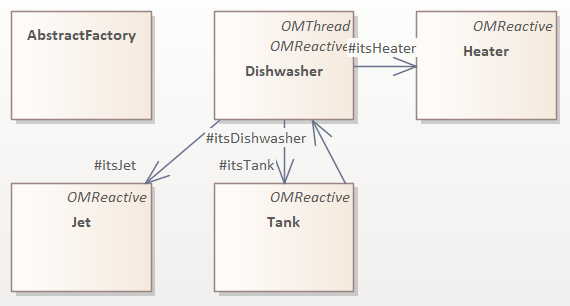}
    \caption{\centering Static View - Component Diagram of the Dishwasher core components after abstraction using LLM}
    \label{fig:dishwasher_class_ea}
\end{figure}


\subsection{Behavioral View: State Machine Generation}

\subsubsection{Prompting Strategy and Example Typology}

Following core component identification, we generated the behavioral view by passing the source code of each component to GPT-4o version 2024-11-20 \cite{gpt4o} along with a structured prompt containing a solution plan.
\textit{Note: For better quality diagrams, refer to our repository.}

\begin{mybox}[colframe=cyan!75!black]{System Prompt}
\begin{lstlisting}[language=HTML, basicstyle=\scriptsize, breaklines=true, columns=fullflexible]
<Role>
You are an expert software engineer.
</Role>
<Goal>
Generate a state machine diagram for the c++ code in plantuml format.
</Goal>
<Solution Plan>
1. Understand the source code
2. Extract the candidate states
3. Provide small and summarized description of each state
4. Extract the transition from one state to another
5. What is the trigger that triggered the transition
6. Review the examples and how their corresponding state machine diagram
7. Construct the state machine diagram for the passed code
8. Review the order of the states based on normal logic and using the examples
</Solution Plan>
!! Important !!: The Controller example is customized to bring how parallel states are present. Do not use it for any other reason.
\end{lstlisting}
\end{mybox}

Each component’s logic was given in a single C++ script file, simplifying the input context.
To enhance the model’s understanding, we applied few-shot prompting \cite{few_shots}, wherein pairs of source code (in form of text) and corresponding state machine diagrams were provided as illustrative examples. Since the ground truth state machine diagrams were provided as images, we pass them as such, however, we scaled them to a maximum width of \textit{800px}. This resolution keep all diagram details readable to the human eye and help in decreasing the token consumption.

\begin{mybox}[colframe=red!75!white]{Examples Prompt (repeat for each example)}
\begin{lstlisting}[basicstyle=\scriptsize, breaklines=true, columns=fullflexible]
Generate a state machine  for the following code:
{source code} + {the ground truth state machine diagram as an image} 
\end{lstlisting}
\end{mybox}

\begin{mybox}[colframe=blue!75!white]{User Prompt}
\begin{lstlisting}[basicstyle=\scriptsize, breaklines=true, columns=fullflexible]
Generate a state machine  for the following code:
{source code for the target class}
\end{lstlisting}
\end{mybox}

\begin{figure}
    \centering
    \includegraphics[scale=0.25]{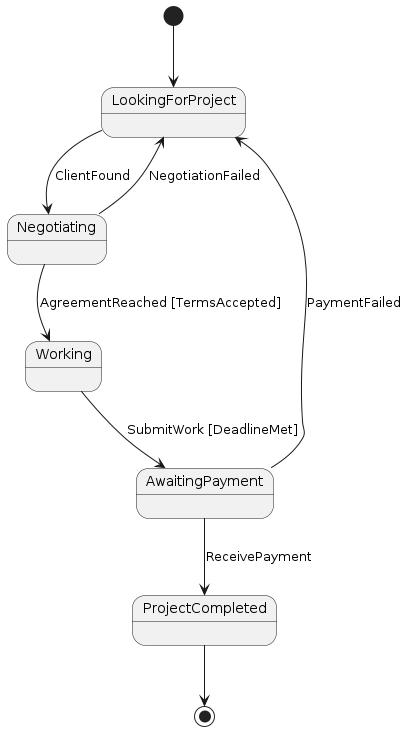}
    \caption{\centering General example - Freelance normal workflow}
    \label{fig:freelance_state}
\end{figure}

We categorized these few-shot examples into three types:
\begin{inparaenum}
    \item \textit{General}: LLM-generated examples with simple behavioral logic: car door (showing the flow between Closed, Open, Locked states), freelance developer (how freelancer accept or decline an order), online retail (how a new order is paid, shipped, delivered, or cancelled flow). Because of the limited space, Fig. \ref{fig:freelance_state} show only the freelance general example, to show how the general examples are constructed.
    \item \textit{Expert}: Cross-example prompting using ground truth diagrams (e.g., Dishwasher states passed to guide Coffee Machine generation) and vice versa.
    \item \textit{Domain}: Prompting within the same domain, where the state machine of one of the core components is generated using the other core components as examples.
\end{inparaenum}

These variations aimed to evaluate how different levels of expertise transfer influence to LLM’s generation quality, where \textit{"Expert"} examples are to simulate how an expert is passing his skill and influencing the output, \textit{"Domain"} the experts' knowledge in addition to his skill, and \textit{"General"} examples are just to assess if the LLM is capable of generating the behavior diagrams without the need for any influence from an expert.

\subsubsection{Evaluation Criteria and Scoring Scheme}

The focus of this paper is to prove a functional semi-automated approach for generating SAD. Accordingly, evaluating the generated state machine diagram is paramount to absolutely assess the functionality of the approach, and it does produce valid outputs.
Thus, we propose a systematical scheme -based on our use case and can be generalized further- to assess the quality of the generated state machine diagrams based on the following key aspects:

\begin{enumerate}
    \item[Q1)] Can the LLM correctly identify the start and end states?
    \item[Q2)] Does the diagram include all ground truth states?
    \item[Q3)] Are the names of the generated states semantically accurate?
    \item[Q4)] Are all reference (ground truth) transitions present in the model?
    \item[Q5)] Are the triggers for each transition correctly inferred?
    \item[Q6)] Is every generated state reachable through at least one path (no dead states)?
    \item[Q7)] Are self-transitions correctly recognized and represented?
    \item[Q8)] Can the model capture parallel or concurrent state behaviors, when applicable?
    \item[Q9)] Does the overall diagram exhibit logical coherence and behavioral consistency in comparison to the ground truth?
\end{enumerate}

Each question was scored as \textit{"X/Y (Z)"}, where \textit{"X"} is the number of matched elements, \textit{"Y"} is the total in the ground truth, and \textit{"Z"} denotes the number of hallucinated elements, that are not found in the ground truth. A perfect match to the ground truth diagram is one where \textit{"X = Y"} and \textit{"Z = 0"}.
For instance: If a generated state machine diagram contains a total of 7 transitions and in the ground truth only 6 ones. Out of the 7 generated transitions, only 5 are matching, and the last two transitions are not present in comparison to the ground truth. Therefor, the score is shown as 5/6 (2).

\subsubsection{Comparative Analysis and Observations}

To ensure consistency and reliability, the state machine diagrams were generated multiple times. The evaluation was conducted manually by comparing the most representative output \footnote{The file name with (picked) in the start of it.} against the ground truth. Based on the defined scoring criteria, the results for the \textit{"CoffeeMachine"} and \textit{"Dishwasher"} components are reported in Tables~\ref{tab:coffee_state_analysis} and \ref{tab:dishwasher_state_analysis}, respectively.

\begin{table*}[!h]
    \centering
    \caption{\centering Coffee Machine State Machine Diagram Scores}
    \label{tab:coffee_state_analysis}
    \rowcolors{2}{white}{blue!10}  
    \renewcommand{\arraystretch}{1.35}  
    \begin{tabular}{c|c|c|c|c|c|c|c|c|c|c}
        \toprule
        Component & Example & Q1 & Q2 & Q3 & Q4 & Q5 & Q6 & Q7 & Q8 & Q9 \\
        \midrule
        \cellcolor{white} \multirow{3}{*}{\vspace{+2.8em}Display} & general & 1/2 (0) & 4/4 (0) & 4/4 (0) & 5/8 (0) & 5/5 (0) & 3/4 (0) & 0/2 (0) & 0/0 (0) & 0/1 (0) \\
         \cellcolor{white} & expert & 1/2 (1) & 4/4 (0) & 4/4 (0) & 7/8 (1) & 5/7 (0) & 3/4 (0) & 2/2 (0) & 0/0 (0) & 1/1 (0) \\
         \cellcolor{white} & domain & 1/2 (1) & 4/4 (0) & 4/4 (0) & 6/8 (1) & 5/6 (0) & 4/4 (0) & 1/2 (0) & 0/0 (0) & 1/1 (0) \\
        \arrayrulecolor{gray}\midrule
        \cellcolor{white} \multirow{3}{*}{\vspace{+2.8em}Boiler} & general & 2/2 (0) & 6/6 (0) & 4/6 (0) & 4/11 (0) & 4/4 (0) & 6/6 (0) & 0/1 (0) & 0/0 (0) & 0/1 (0) \\
         \cellcolor{white} & expert & 2/2 (0) & 6/6 (1) & 4/6 (1) & 10/11 (0) & 8/10 (0) & 6/6 (0) & 0/1 (0) & 0/0 (0) & 1/1 (0) \\
         \cellcolor{white} & domain & 2/2 (0) & 6/6 (1) & 4/6 (1) & 10/11 (0) & 8/10 (0) & 6/6 (0) & 0/1 (0) & 0/0 (0) & 1/1 (0) \\
        \arrayrulecolor{gray}\midrule
        \cellcolor{white} \multirow{3}{*}{\vspace{+2.8em}CoffeeMachine} & general & 4/5 (0) & 12/16 (0) & 12/12 (0) & 15/26 (0) & 7/15 (0) & 12/12 (0) & 0/7 (0) & 0/2 (0) & 0/1 (0) \\
         \cellcolor{white} & expert & 2/5 (0) & 12/16 (1) & 12/12 (1) & 13/26 (2) & 3/13 (0) & 10/12 (0) & 0/7 (0) & 0/2 (0) & 0/1 (0) \\
         \cellcolor{white} & domain & 3/5 (0) & 16/16 (3) & 13/16 (6) & 14/26 (3) & 14/14 (0) & 14/16 (0) & 0/7 (0) & 1/2 (0) & 0/1 (0) \\
        \bottomrule
    \end{tabular}
\end{table*}

\begin{table*}[!h]
    \centering
    \caption{\centering Dishwasher State Machine Diagram Scores}
    \label{tab:dishwasher_state_analysis}
    \rowcolors{2}{white}{gray!20} 
    \renewcommand{\arraystretch}{1.35} 
    \begin{tabular}{c|c|c|c|c|c|c|c|c|c|c}
        \toprule
        Component & Example & Q1 & Q2 & Q3 & Q4 & Q5 & Q6 & Q7 & Q8 & Q9 \\
        \midrule
        \cellcolor{white}\multirow{3}{*}{\vspace{+2.8em}Heater} & general & 1/1 (0) & 2/2 (0) & 2/2 (0) & 3/3 (0) & 3/3 (0) & 2/2 (0) & 0/0 (0) & 0/0 (0) & 1/1 (0) \\
        \cellcolor{white} & expert & 1/1 (0) & 2/2 (0) & 2/2 (0) & 3/3 (0) & 3/3 (0) & 2/2 (0) & 0/0 (0) & 0/0 (0) & 1/1 (0) \\
        \cellcolor{white} & domain & 1/1 (0) & 2/2 (0) & 2/2 (0) & 3/3 (0) & 3/3 (0) & 2/2 (0) & 0/0 (0) & 0/0 (0) & 1/1 (0) \\
        \arrayrulecolor{gray}\midrule
        \cellcolor{white}\multirow{3}{*}{\vspace{+2.8em}Jet} & general & 1/2 (0) & 3/4 (0) & 3/3 (0) & 5/5 (0) & 5/5 (0) & 3/3 (0) & 0/0 (0) & 0/0 (0) & 1/1 (0) \\
        \cellcolor{white} & expert & 1/2 (0) & 4/4 (0) & 4/4 (0) & 7/5 (0) & 7/7 (0) & 4/4 (0) & 0/0 (0) & 0/0 (0) & 1/1 (0) \\
        \cellcolor{white} & domain & 1/2 (0) & 3/4 (0) & 3/3 (0) & 5/5 (0) & 5/5 (0) & 3/3 (0) & 0/0 (0) & 0/0 (0) & 1/1 (0) \\
        \arrayrulecolor{gray}\midrule
        \cellcolor{white}\multirow{3}{*}{\vspace{+2.8em}Tank} & general & 1/1 (0) & 4/4 (0) & 4/4 (0) & 5/5 (3) & 5/5 (0) & 4/4 (0) & 0/0 (2) & 0/0 (0) & 0/1 (0) \\
        \cellcolor{white} & expert & 1/1 (1) & 4/4 (0) & 4/4 (0) & 5/5 (2) & 5/5 (0) & 4/4 (0) & 0/0 (0) & 0/0 (0) & 0/1 (0) \\
        \cellcolor{white} & domain & 1/1 (0) & 4/4 (0) & 4/4 (0) & 5/5 (2) & 5/5 (0) & 4/4 (0) & 0/0 (0) & 0/0 (0) & 0/1 (0) \\
        \arrayrulecolor{gray}\midrule
        \cellcolor{white}\multirow{3}{*}{\vspace{+2.8em}Dishwasher} & general & 1/5 (0) & 8/15 (0) & 8/8 (0) & 8/24 (0) & 4/8 (0) & 8/8 (0) & 0/2 (0) & 0/4 (0) & 0/1 (0) \\
        \cellcolor{white} & expert & 1/5 (0) & 8/15 (0) & 8/8 (0) & 8/24 (2) & 8/8 (0) & 8/8 (0) & 0/2 (0) & 0/4 (0) & 0/1 (0) \\
        \cellcolor{white} & domain & 4/5 (0) & 14/15 (0) & 13/14 (1) & 12/24 (0) & 12/12 (0) & 14/14 (0) & 0/2 (0) & 3/4 (0) & 0/1 (0) \\
        \bottomrule
    \end{tabular}
\end{table*}


We observed that smaller, less complex classes tend to yield higher-fidelity diagrams. In particular, classes such as \textit{"Display"} and \textit{"Boiler"} regarding the \textit{"CoffeeMachine"}, and \textit{"Heater"}, \textit{"Jet"}, and \textit{"Tank"} regarding the \textit{"Dishwasher"}, exhibited strong alignment with the ground truth in terms of state coverage, transitions, triggers, and naming accuracy.

\begin{figure}
    \centering
    \includegraphics[width=\columnwidth]{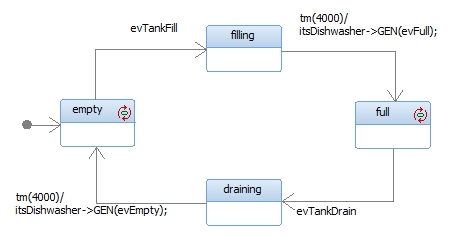}
    \caption{\centering Ground truth state machine diagram for the Tank component from Rhapsody}
    \label{fig:tank_gt}
\end{figure}

\begin{figure}
    \centering
    \includegraphics[scale=0.35]{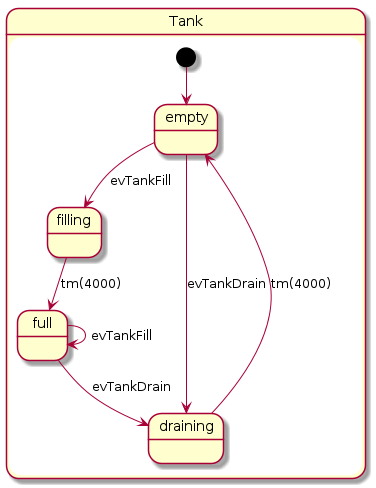}
    \caption{\centering Generated state machine diagram for the Tank component using GPT-4o and Domain Examples}
    \label{fig:tank_gen}
\end{figure}

\begin{figure}
    \centering
    \includegraphics[width=\columnwidth]{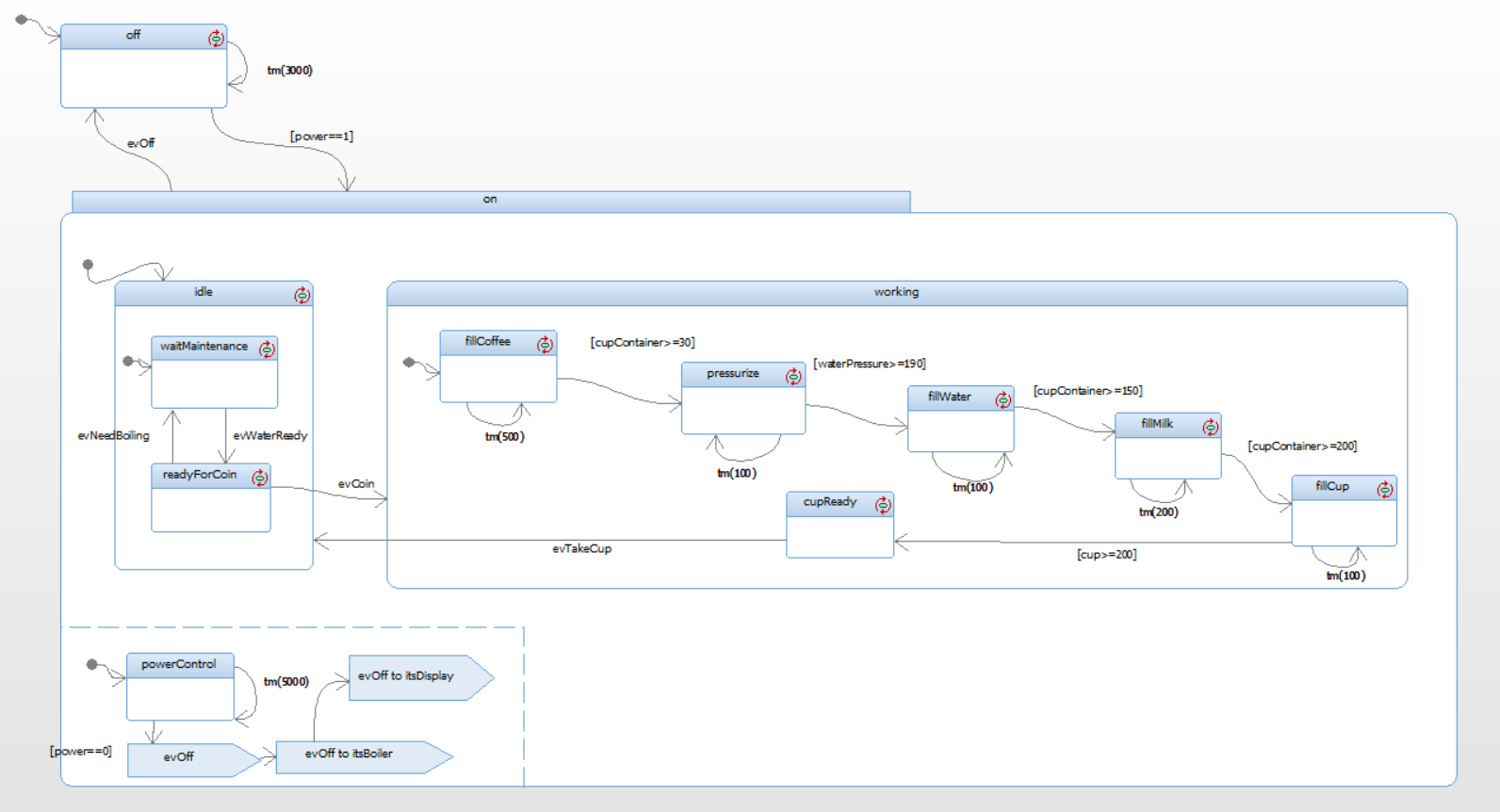}
    \caption{\centering Ground truth state machine diagram for the Coffee Machine component from Rhapsody}
    \label{fig:coffe_state_gt}
\end{figure}

\begin{figure}
    \centering
    \includegraphics[width=\columnwidth]{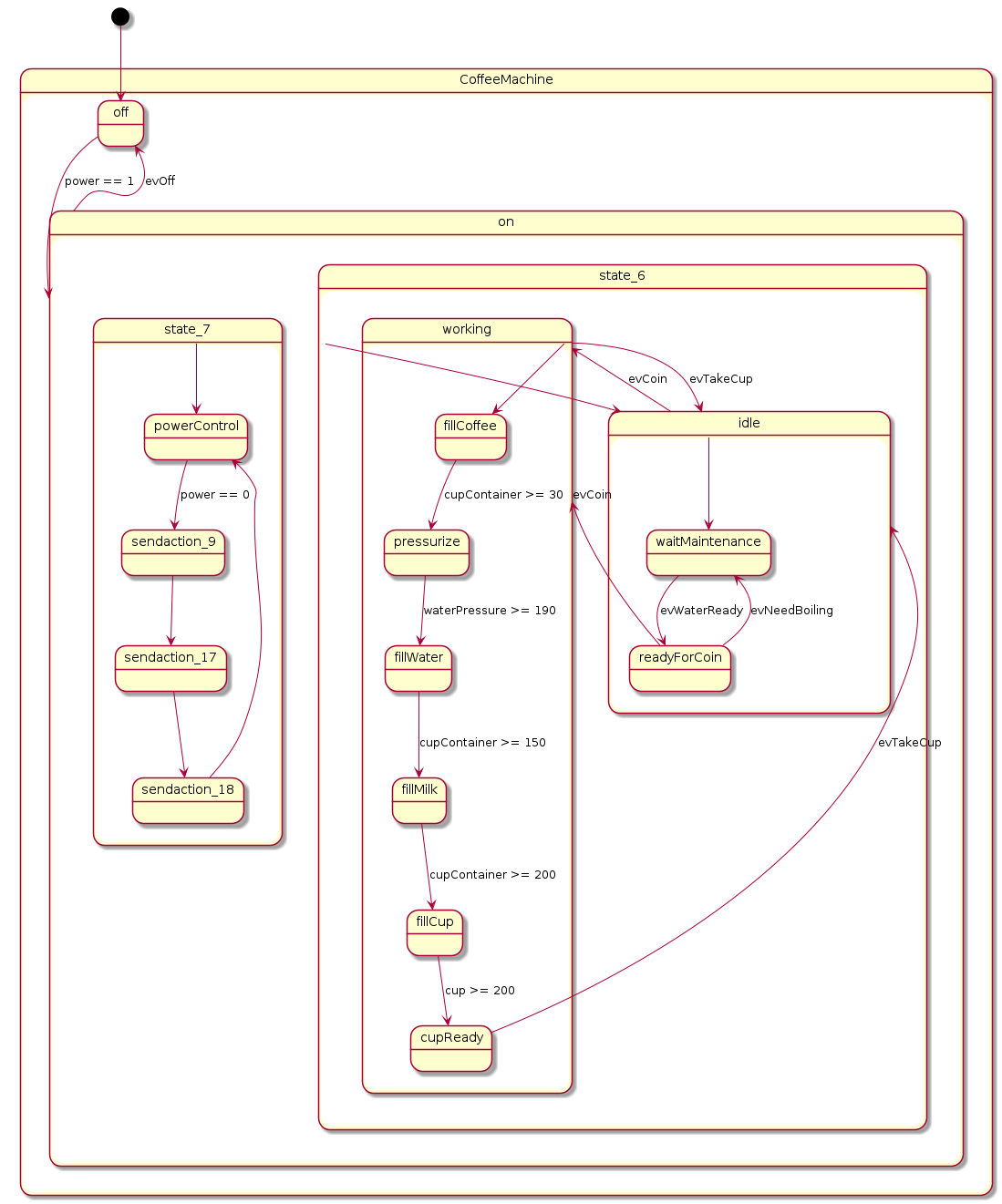}
    \caption{\centering Generated state machine diagram for the Coffee Machine component using GPT-4o and Domain Examples}
    \label{fig:coffee_state_gen}
\end{figure}

However, only \textit{"Heater"}, \textit{"Jet"}, and \textit{"Boiler"} were acceptable diagrams when using general examples.
Other classes, despite appearing close in numeric score, exhibited critical absence of the start states within substates in some cases and in another cases a hallucinated transition. These factors disrupted the flow and rendered the generated models behaviorally incomplete.
For instance, as shown in Fig.\ref{fig:coffe_state_gt}, the ground truth for the \textit{"CoffeeMachine"} involves a transition from \textit{"off"} to \textit{"on"}, followed by two parallel states. In the corresponding generated version (Fig.\ref{fig:coffee_state_gen}), although the transition from \textit{"off"} to \textit{"on"} is captured, the entry point within the \textit{"on"} composite state is missing, undermining the execution flow.
Another one is \textit{"Tank"} component where a hallucinated transition from \textit{"empty"} to \textit{"draining"} and the self transition in the state \textit{"full"} corrupted the flow as shown in ground truth Fig.\ref{fig:tank_gt} and the generated diagram Fig.\ref{fig:tank_gen}.


\begin{figure}
    \centering
    \includegraphics[width=\columnwidth]{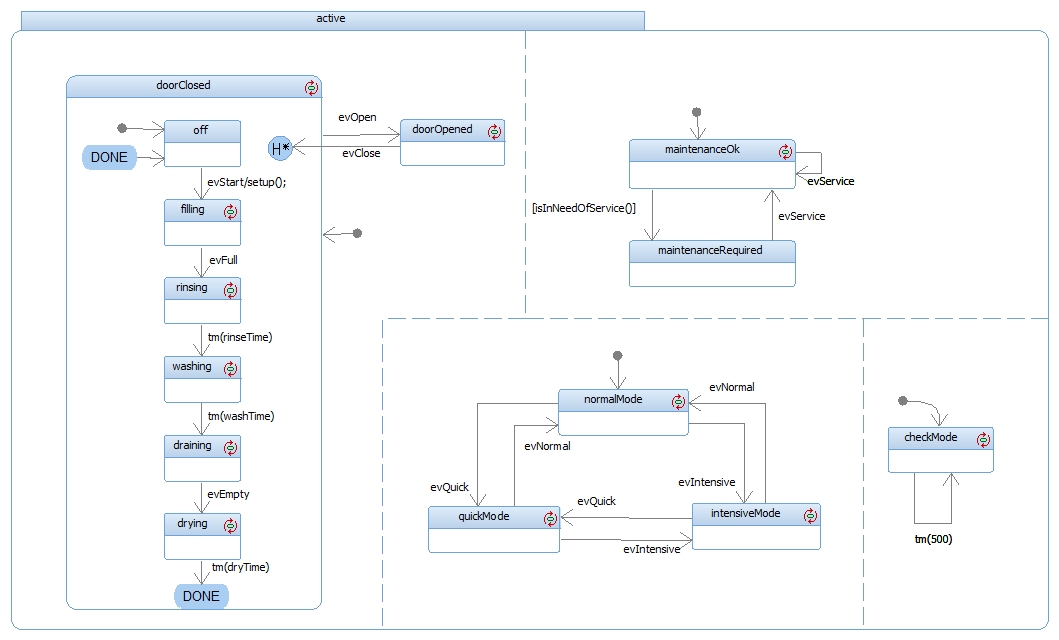}
    \caption{\centering Ground truth state machine diagram for the Dishwasher component from Rhapsody}
    \label{fig:dish_state_gt}
\end{figure}

\begin{figure}
    \centering
    \includegraphics[scale=0.27]{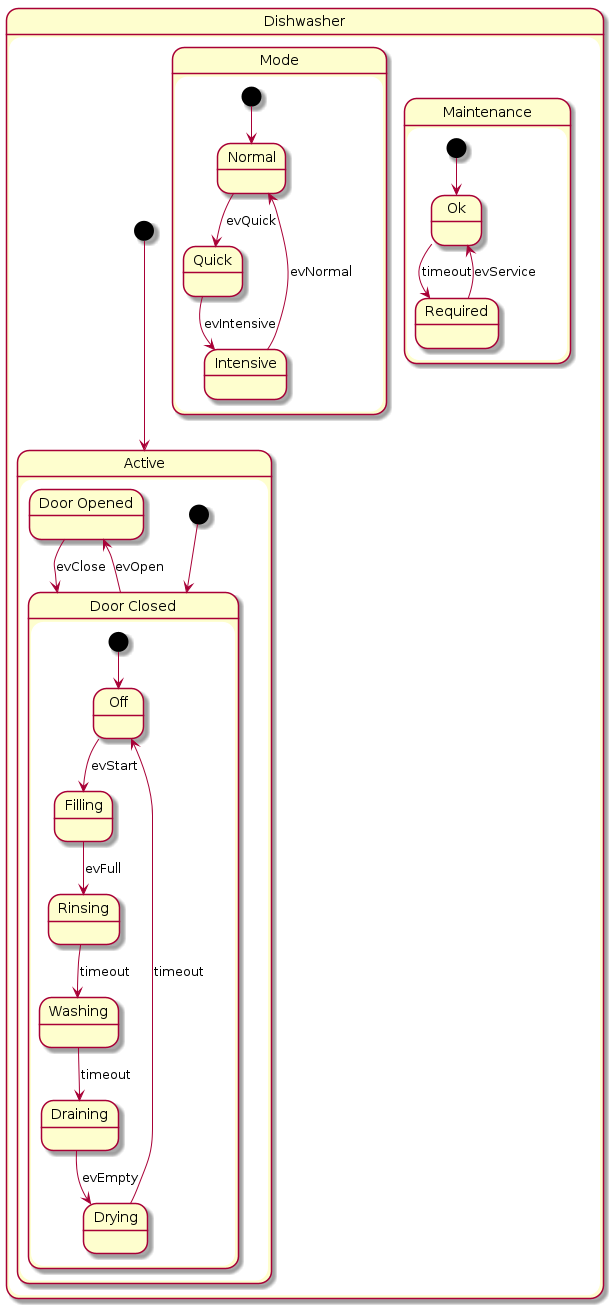}
    \caption{\centering Generated state machine diagram for the Dishwasher component using GPT-4o and Domain Examples}
    \label{fig:dish_state_gen}
\end{figure}

The two most complex component—\textit{"CoffeeMachine"} and \textit{"Dishwasher"}—presented the most challenges but also offered the richest insights.
Their ground truth are shown in Figs.~\ref{fig:coffe_state_gt}, \ref{fig:dish_state_gt}, and generated diagrams are in, \ref{fig:coffee_state_gen}, and \ref{fig:dish_state_gen} respectively.
Due to space limitations, not all diagrams are shown. We refer readers to our GitHub repository for full visual references \cite{gh_repo}.


In answering the evaluation questions, several patterns emerged:

\begin{inparaenum}[1) ]
    \item Start states within substates were frequently missing, breaking the logical flow and violating the sequential execution expected in state hierarchies.

    \item The generated diagrams sometimes adopted a hierarchical structure and other times a flat one. This variability means that what may appear as hallucinated states were, in fact, structural necessities to support hierarchy.

    \item State names generally reflected the terminology used in source code, indicating that the LLM leverages lexical cues from the implementation.

    \item Transitions were largely captured, but self-transitions—often associated with timeouts—were almost entirely absent.

    \item Timeout transitions were inconsistently labeled, frequently missing the actual time duration and instead annotated generically as \textit{timeout}.

    \item Parallel states were not reliably identified in general or expert examples. However, domain-specific examples substantially improved recognition of concurrent behaviors.

    \item Certain state types, such as the \textit{"H*"} pseudostate and final state \textit{"Done"} in the \textit{Dishwasher} class, were entirely omitted.

    \item As a consequence of these issues—particularly the missing substate entry points—the overall behavioral logic for the largest classes was deemed unsatisfactory, despite domain-specific prompting notably improved the LLM generation quality.
\end{inparaenum}

Different example types used during generation yielded varied levels of quality. Diagrams generated using domain-specific examples showed the strongest alignment with ground truth, successfully modeling complex behaviors such as parallelism. In contrast, expert examples were adequate only for less complex cases, while general examples performed poorly in all dimensions other than the simple straight forward ones.

These findings suggest that while GPT-4o \cite{gpt4o} possesses a foundational ability to infer behavioral models from code, this capability is highly sensitive to the structure and semantic relevance of the input examples. Specifically, incorporating domain-aligned few-shot prompts is essential for achieving accurate, usable behavioral diagrams in real-world projects.


\section{Conclusion and Future Work} \label{conclusion}

In this work, we proposed a hybrid approach for semi-automatically generating SAD from source code by combining RE with LLM-guided abstraction.
The method effectively recovered both static and behavioral views, with the LLM identifying core components and generating meaningful state machine diagrams with the help of the proposed systematic assessment.

Reliance on LLM-guided abstraction for core component detection showed highly promising usability.
However, the generalizability remains an open question.

Moreover, the generation of the state machine diagram using domain-specific examples notably improved behavioral modeling, implicitly injecting expert knowledge into the process.
The pre-trained LLM is also continuously trained on base knowledge that is improving over time and might overcome the need for high-quality domain examples.

Nonetheless, a consistent challenge is the context window; some components involve code that exceeds the context windows, requiring a technique for maintaining contextual relevance for the LLM to correctly infer the component's behavior, such as iteratively prompting the source code to build the state machine diagram step by step.

Addressing the research questions, our approach showed the ability to semi-automate the process of generating static and behavioral diagrams, where the RE step is manually done and the LLM is automated.
Moreover, the LLM with the help of the domain examples successfully inferred the code behavior and translated it to a state machine diagram.
However, the high-complexity diagrams were occasionally logically flawed, and the less complex ones yielded superior, more human-like diagrams.

Further work should also focus on improving behavioral inference; promising directions include using LLM agents \cite{agent} and integrating reasoning-capable models like DeepSeek \cite{r1} or Phi-4 \cite{phi4} to better capture implicit logic, state hierarchies, and parallel states.
Such enhancements aim to increase the precision, flexibility, and autonomy of automated SAD generation.


\section*{Acknowledgment}
The authors extend their sincere gratitude to Dr. Damian Hofmann, Dr. Christoph Raab, and Dr. Christian Nabert for their invaluable and insightful technical discussions, which significantly enhanced the clarity and validity of the reported findings.
Furthermore, the authors acknowledge the use of generative Artificial Intelligence tools, specifically ChatGPT and Google Gemini, for non-substantive editorial assistance in refining the manuscript's language, grammar, and structural flow.


\bibliographystyle{IEEEtran}
\bibliography{bib}


\end{document}